\documentclass[3p,times,twocolumn]{elsarticle}

\usepackage{ecrc}

\volume{00}

\firstpage{1}

\journalname{Nuclear and Particle Physics Proceedings}

\runauth{}

\jid{nppp}

\jnltitlelogo{Nuclear and Particle Physics Proceedings}

\usepackage{amssymb}

\usepackage[figuresright]{rotating}

\begin{document}

\begin{frontmatter}



\dochead{}

\title{Holographic Photon Production and Anisotropic Flow}


\author{Ioannis Iatrakis$^{1}$,
	Elias Kiritsis$^{2,3,4}$,	Chun Shen$^{5,6}$, 
	Di-Lun Yang$^{7}$}

\address{$^{1}$Institute for Theoretical Physics and Center for Extreme Matter and Emergent Phenomena, Utrecht University, Leuvenlaan 4, 3584 CE Utrecht, The Netherlands.\\
$^2$Crete Center for Theoretical Physics, Institute of Theoretical and Computational Physics, Department of Physics University of Crete, 71003 Heraklion, Greece.\\
$^3$ Crete Center for Quantum Complexity and Nanotechnology, Department of Physics, University of Crete, 71003 Heraklion, Greece.\\
$^4$APC, Univ Paris Diderot, Sorbonne Paris Cit\'e, APC, UMR 7164 CNRS, F-75205 Paris, France.\\
$^5$Department of Physics, McGill University, 3600 University Street Montreal QC, H3A 2T8, Canada.\\
$^6$Physics Department, Brookhaven National Laboratory, Upton, NY 11973, USA.
\\
$^7$Theoretical Research Division, Nishina Center, RIKEN, Wako, Saitama 351-0198, Japan.
	 }
\begin{abstract}
The thermal-photon emission from strongly coupled gauge theories at finite temperature via the bottom-up models in holographic QCD in the deconfined phase is studied. The models are constructed to approximately reproduce the electric conductivity obtained from lattice simulations for the quark gluon plasma (QGP). The emission rates are then embedded in hydrodynamic simulations combined with prompt photons and hadronic contributions to analyze the spectra and anisotropic flow of direct photons in RHIC and LHC. In general, the holographic models enhance the yield and improve the agreement in spectra, while they reduce the flow in low $p_T$ and increase it in high $p_T$. 
\end{abstract}

\end{frontmatter}

\begin{flushleft}
	CCQCN-2016-176 CCTP-2016-18
\end{flushleft}

\section{Introduction}
\label{}
Electromagnetic probes such as photons and dileptons play important roles in both relativistic heavy ion collisions and cosmology. In heavy ion collisions, considerable amount of direct photons, which do not come from hadron decays, is emitted by the thermal photons emitted from the quark gluon plasma (QGP).
Particularly, recent measurements of large anisotropic flow of direct photons comparable to the hadron flow in RHIC \cite{Adare:2015lcd} and LHC \cite{Lohner:2012ct} lead to a new puzzle. In general, electromagnetic probes barely interact with the QGP after they are produced and only record the local information in heavy ion collisions, although the scenario could be modified for long-lived plasmas such as the cosmic plasma \cite{Muller:2015maa,Yang:2015bva}. On the other hand, from most of theoretical predictions, the direct-photon spectra are underestimated as well. 
There have been intensive studies to reconcile the tension between experimental observations and theoretical predictions \cite{Shen:2013cca,Paquet:2015lta}. Furthermore, some of novel mechanisms such as the enhanced thermal-photon production from strong magnetic fields were considered \cite{Basar:2012bp,Muller:2013ila}.  

To study thermal-photon emission from the QGP, in perturbative quantum chromodynamics (pQCD), the complete leading-order result of photon production in thermal equilibrium incorporating 2 to 2 scattering and collinear emission has been obtained in \cite{Arnold:2001ms}.
However, due to the strong coupling of the QGP around the deconfinement temperature, perturbative approaches may become invalid.
Alternatively, the gauge/gravity duality, which connects the strongly coupled gauge theories and classical supergravity, provides a potential way to tackle non-perturbative problems \cite{Maldacena:1997re}. In holography, the study of thermal-photon production and dilepton production from a strongly coupled $\mathcal{N}=4$ SYM plasma was initiated by \cite{CaronHuot:2006te}. In \cite{Iatrakis:2016ugz}, two bottom-up holographic models are applied to model the sQGP, both of which break conformal invariance and match several  properties of QCD at finite temperature as calculated from lattice simulations. The photon-emission rates from these holographic models are then convoluted with the medium evolution. Furthermore, the contributions from prompt photons and thermal photons from hadron gas are incorporated to compute both the spectra and flow of direct photons in both RHIC and LHC energies.
In this proceeding, we review the major results found in \cite{Iatrakis:2016ugz}. 

\section{Direct Photon Spectra}
One of the models employed in \cite{Iatrakis:2016ugz} was introduced by Gubser and Nellore (GN) \cite{Gubser:2008ny}. The other model is V-QCD, which is a more sophisticated holographic model and it is based on the improved holographic QCD formalism plus flavor degrees of freedom \cite{Iatrakis:2010jb,Iatrakis:2010zf}. Both models were constructed to fit the electric conductivity from lattice simulations \cite{Aarts:2014nba}. For V-QCD, two kinds of setup were considered, where one leads to saturation and another one results in monotonic increase of the electric conductivity at high temperature beyond the range of lattice results. The former and the latter are dubbed as VQCD1 and VQCD2, respectively. In general, the corresponding photon-emission rates from holography yield blue-shift, where the maximum shifts to higher momenta, compared to weakly coupled results. One may refer to \cite{Iatrakis:2016ugz} for detailed comparisons of emission rates between different models. 

The holographic models discussed above are only responsible for thermal photons in the QGP phase. To make direct comparisons with experimental data, one has to be embedded them into the medium evolution and include both prompt and thermal photons. More details of the model setup are discussed in \cite{Iatrakis:2016ugz}. Here we highlight the important findings therein.
\begin{figure}[h!]
	\centering
	\begin{tabular}{cc}
		\includegraphics[width=0.5\linewidth,height=0.6\linewidth]{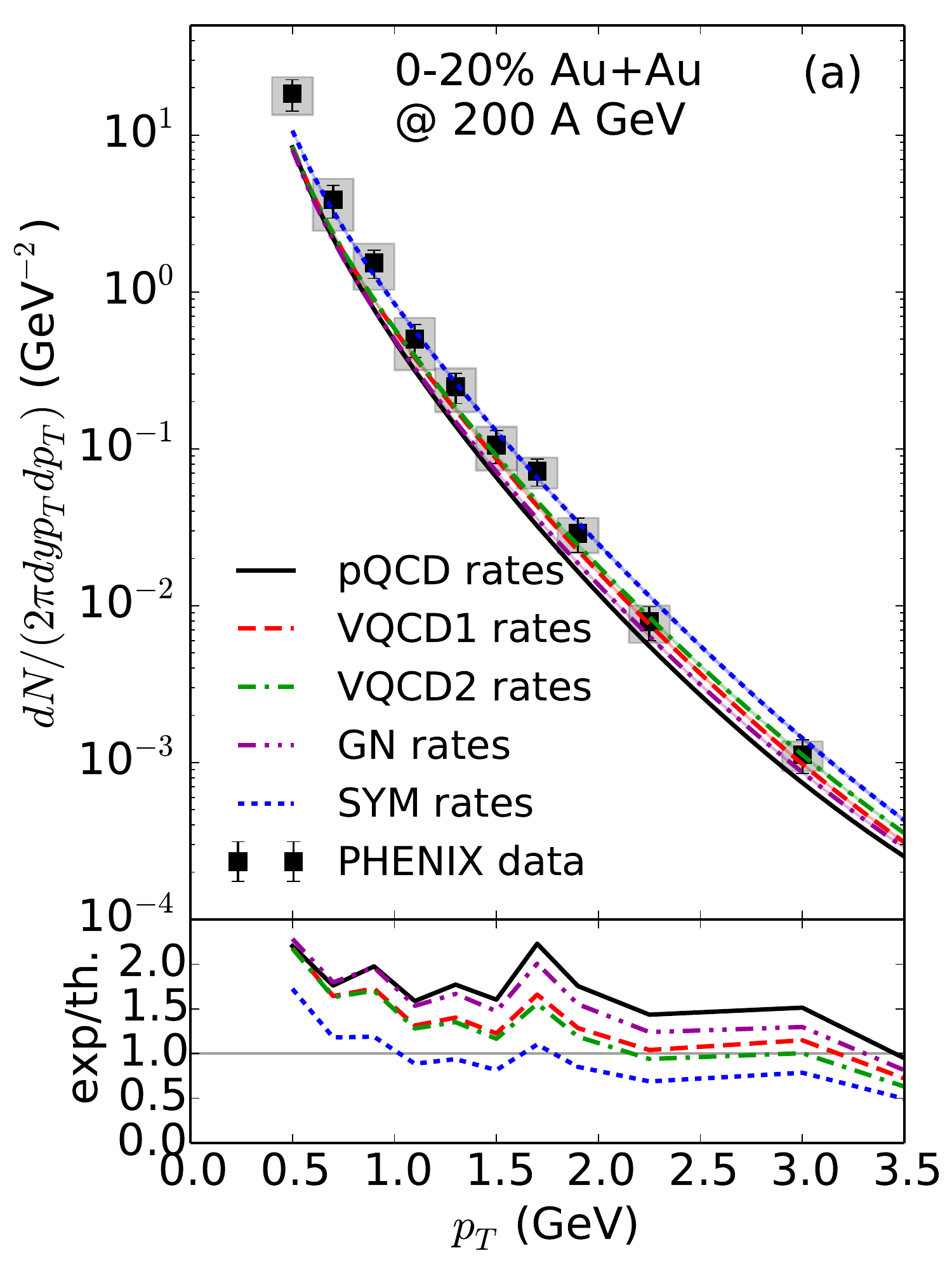} &
		\includegraphics[width=0.5\linewidth,height=0.6\linewidth]{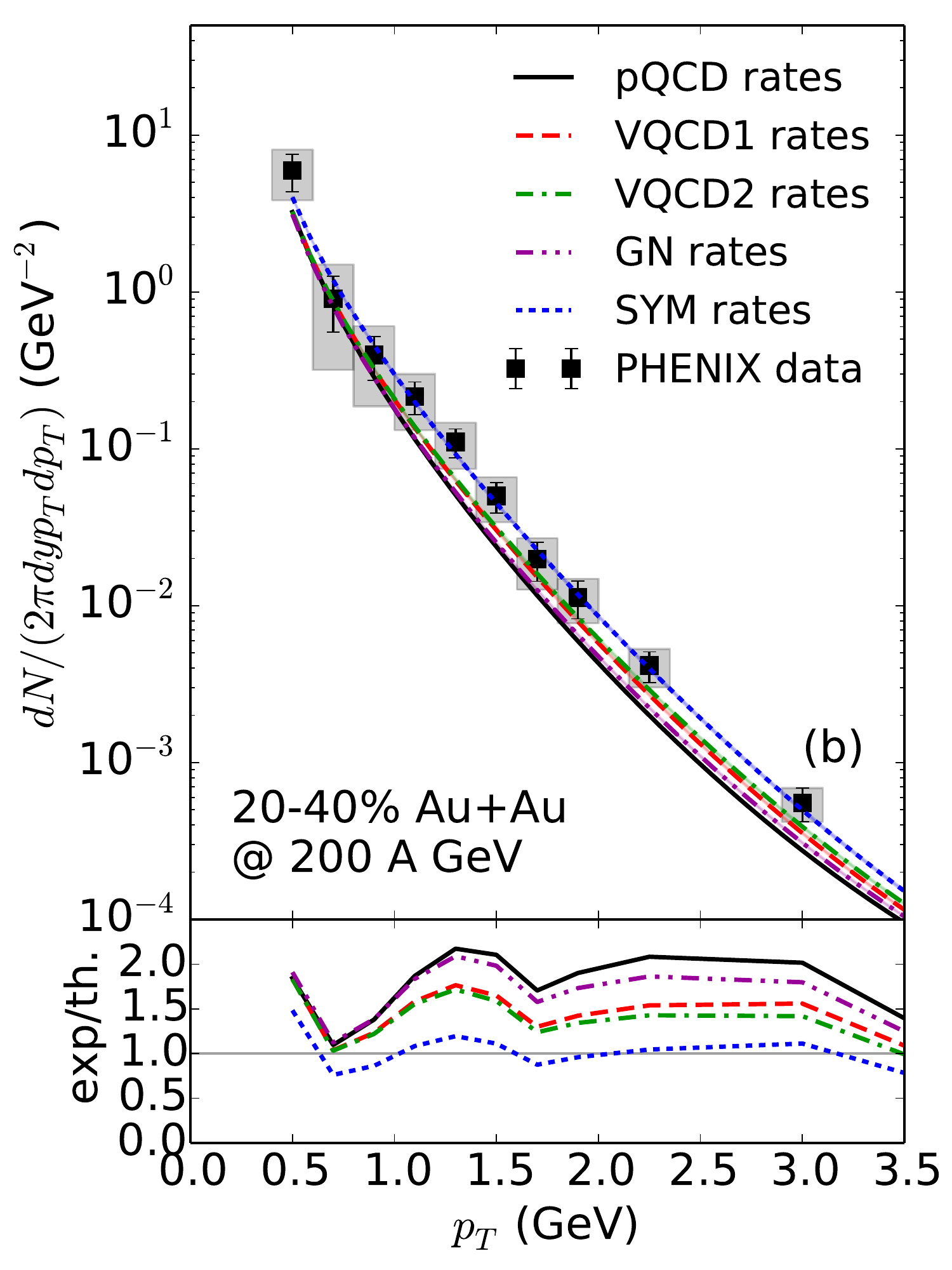} \\
		\includegraphics[width=0.5\linewidth,height=0.6\linewidth]{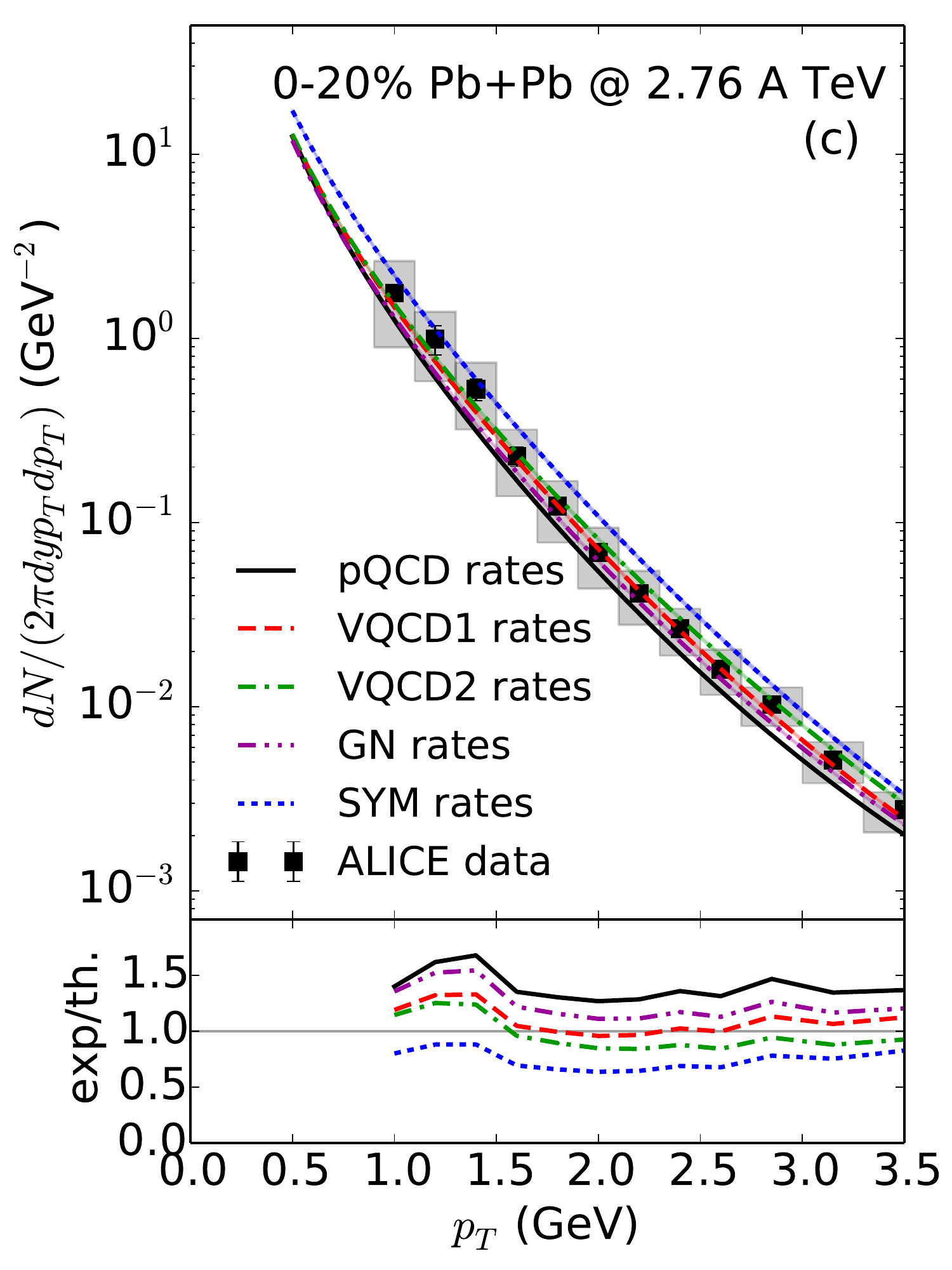} &
		\includegraphics[width=0.5\linewidth,height=0.6\linewidth]{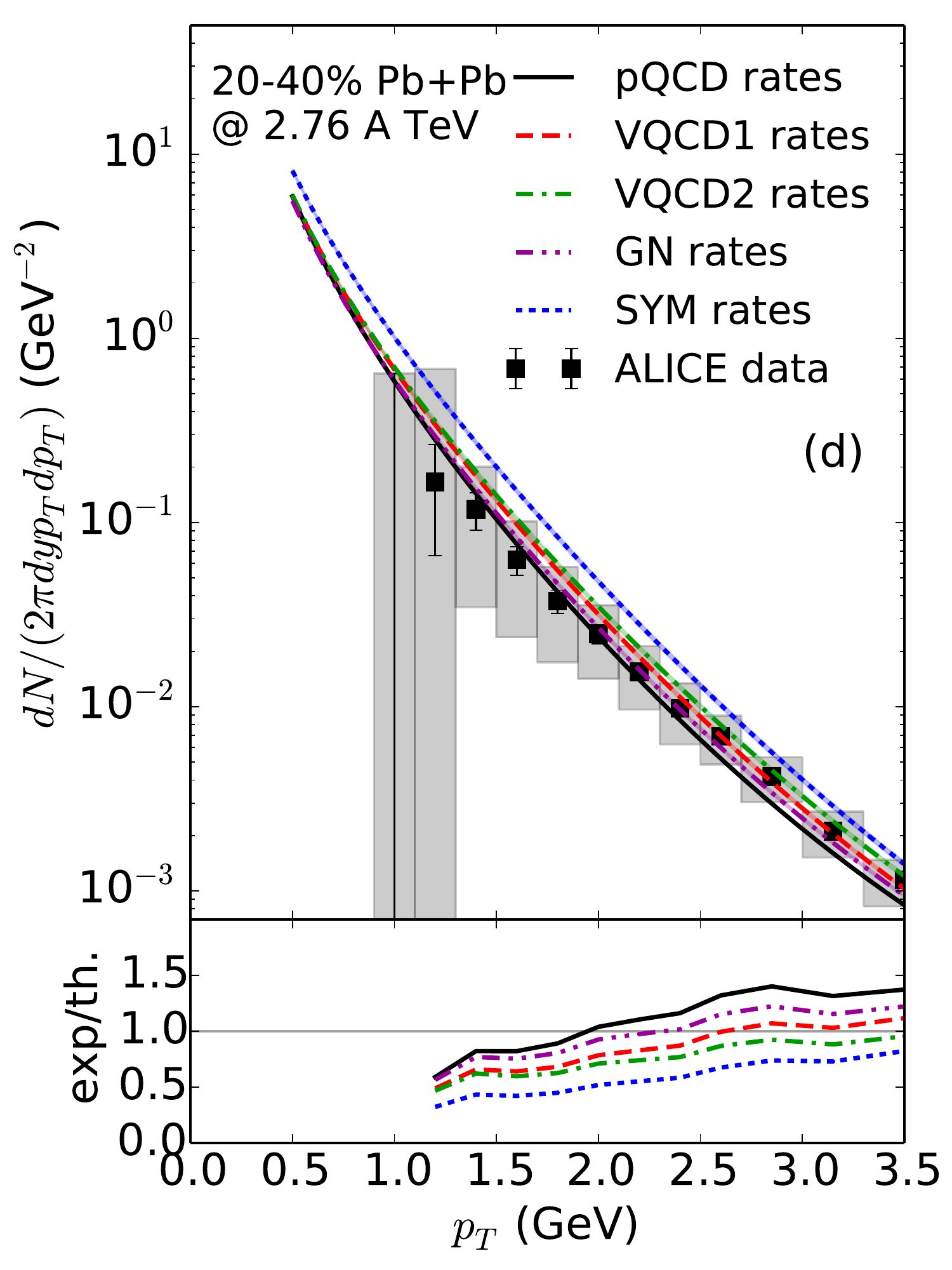}
	\end{tabular}
	\caption{Results for direct-photon spectra \cite{Iatrakis:2016ugz}. (Color online) Direct photon spectra from 0-20\% (a) and 20-40\% (b) Au+Au collisions at 200 GeV compared with the PHENIX measurements \cite{Adare:2014fwh} and from 0-20\% (c) and 20-40\% (d) Pb+Pb collisions at 2.76 $A$\,TeV compared with the ALICE measurements \cite{Adam:2015lda}. The ratios of experimental data to theoretical results are shown in the bottom of each plot.}
	\label{fig6.1}
\end{figure}

Direct photon spectra using different sets of QGP photon rates are compared to the experimental measurements in Au+Au collisions at 200 $A$\,GeV at the RHIC and in Pb+Pb collisions at 2.76 $A$\,TeV at the LHC in Figs.~\ref{fig6.1}.
The QGP photon rates from holographic models result in more thermal radiation compared to the results with the pQCD rate. 
The reasons for this depend on the holographic model.
First, at strong coupling we expect more photon emissions than at weak coupling. On top of this, SYM which contains extra supersymmetric partners is expected to give the highest rate and this is turns out to be correct. The GN and VQCD models are non supersymmetric and have the same number of (perturbative) degrees of freedom as QCD. 

Therefore, among the different holographic rates, the SYM rates give the most direct photons. In heavy ion collisions, most of the thermal photon radiations are coming from the phase transition region, $150 < T < 250$ MeV, where the space-time volume is the largest. In this temperature region, the photon emission rates are suppressed in the VQCD models compared to the SYM rates. On the other hand, the GN model leads to smaller spectra compared to the ones for VQCD and SYM models as expected from the electric conductivity and emission rates. In general, VQCD models somewhat improve the agreement in direct-photon spectra. 

\section{Direct-Photon Flow}
On the one hand, the absolute yield of direct photon spectra provides  information about the system's space-time volume as well as the degrees of freedom of photon emitters in the medium. On the other hand, the anisotropic flows of direct photons are more sensitive to the relative temperature dependence of photon rates and their interplay with the development of hydrodynamic anisotropic flows during the evolution.

\begin{figure}[h!]
	\centering
	\begin{tabular}{cc}
		\includegraphics[width=0.48\linewidth]{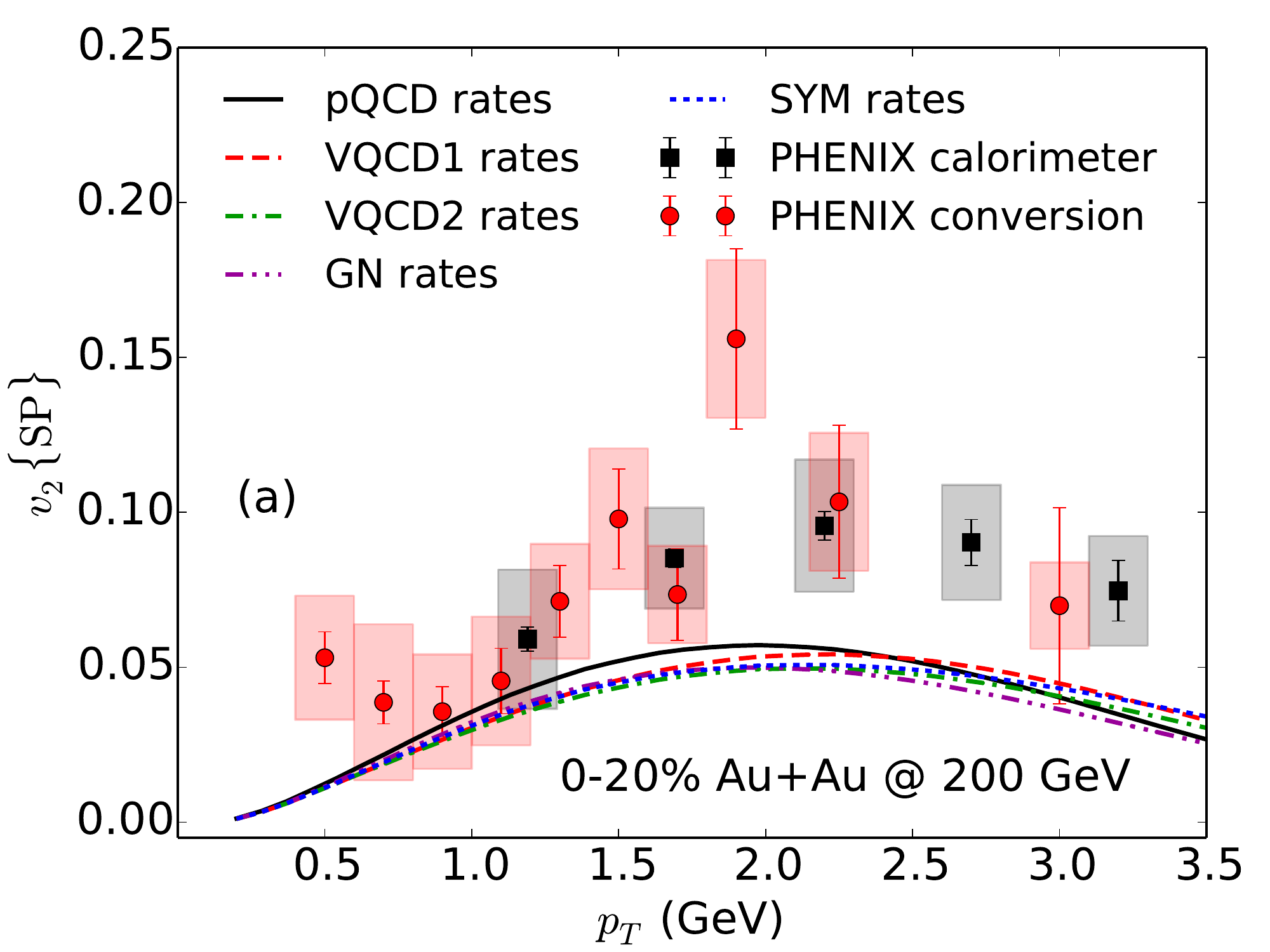} &
		\includegraphics[width=0.48\linewidth]{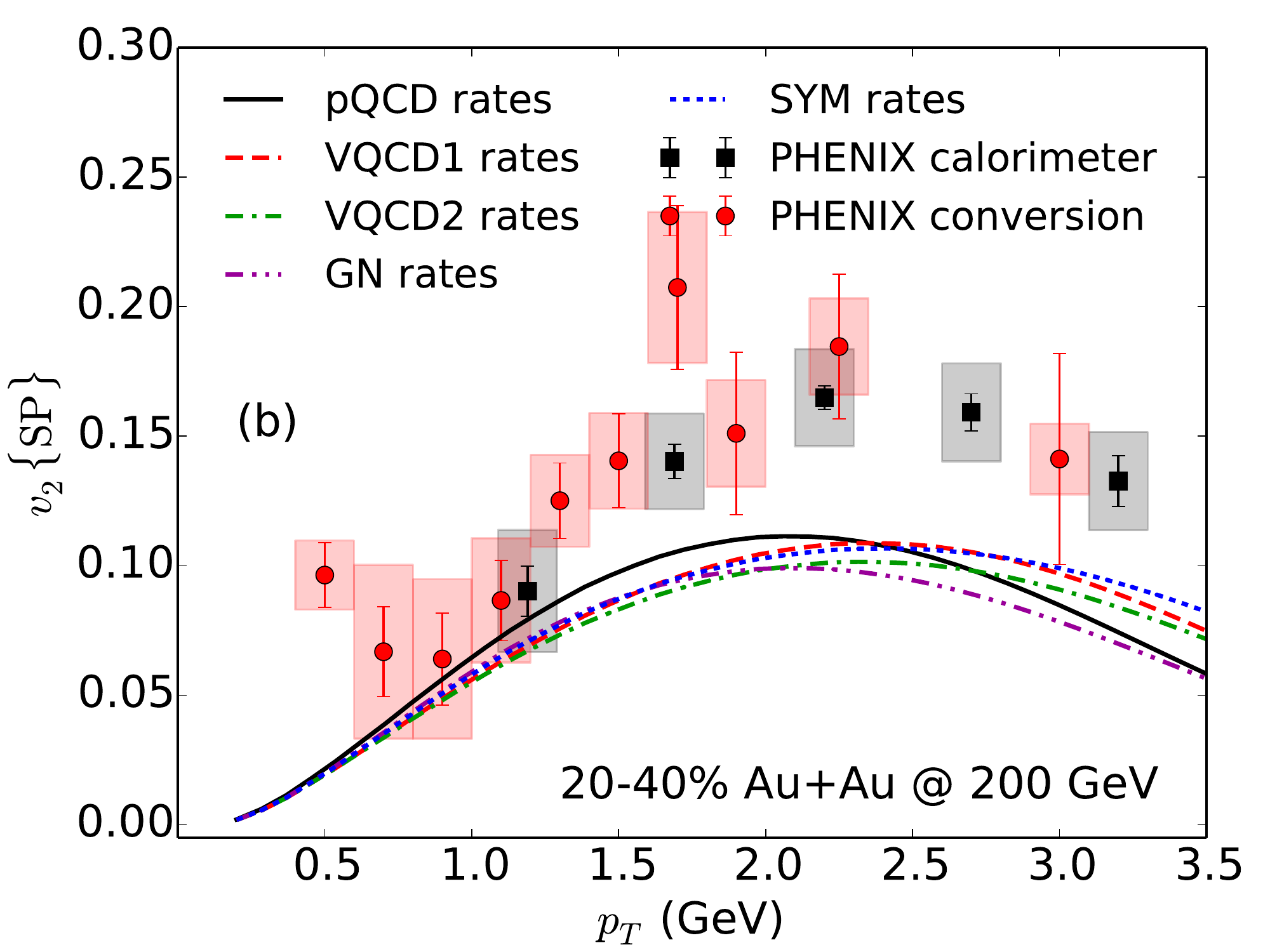} \\
		\includegraphics[width=0.48\linewidth]{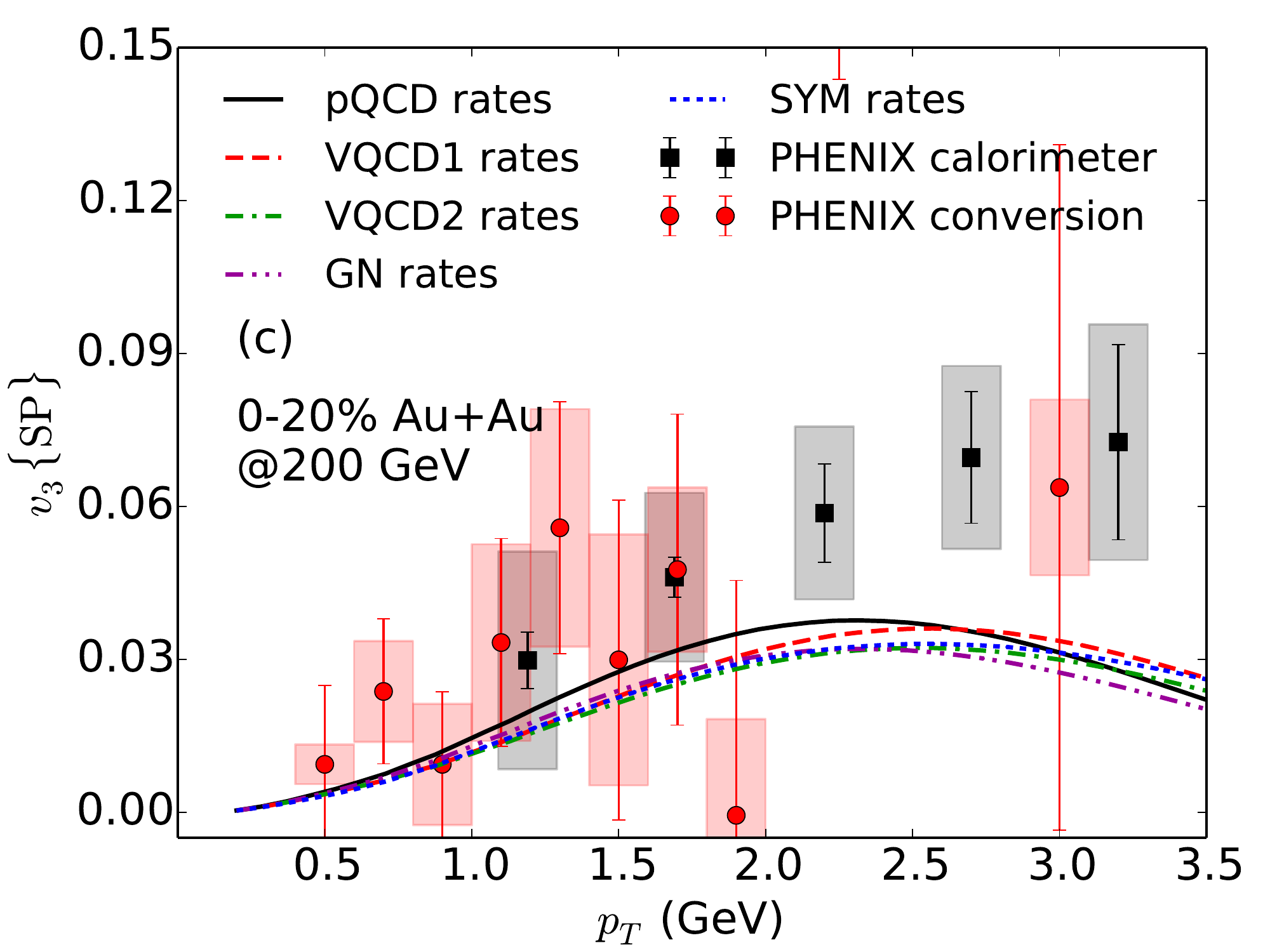} &
		\includegraphics[width=0.48\linewidth]{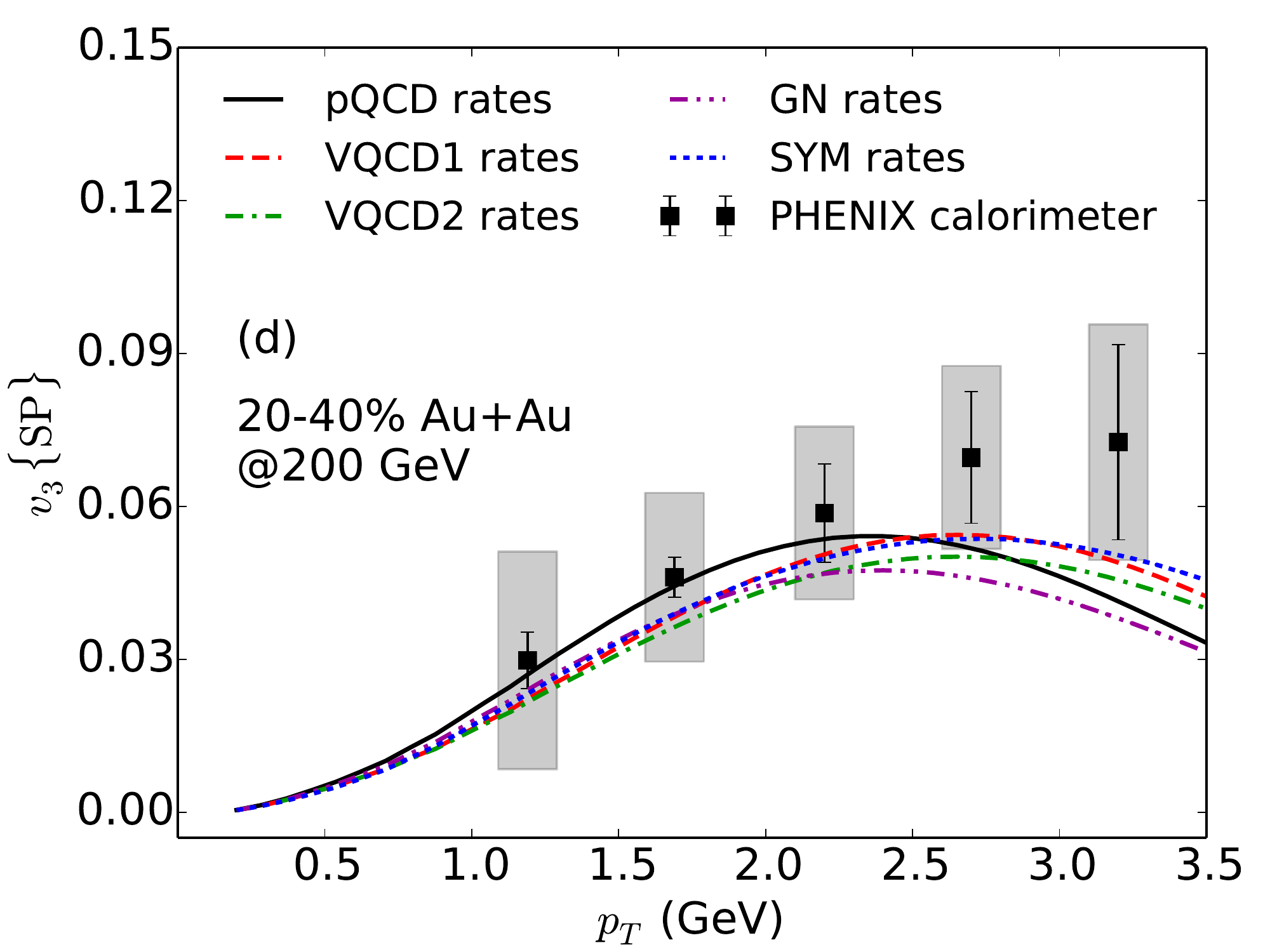}
	\end{tabular}
	\caption{Results for anisotropic flow in RHIC \cite{Iatrakis:2016ugz}. (Color online) Direct photon anisotropic flow $v_{2,3}$ from 0-20\% (a,c) and 20-40\% (b,d) Au+Au collisions at 200 GeV compared with the PHENIX measurements\cite{Adare:2015lcd}.}
	\label{fig6.2}
\end{figure}

\begin{figure}[h!]
	\centering
	\includegraphics[width=0.7\linewidth,height=0.5\linewidth]{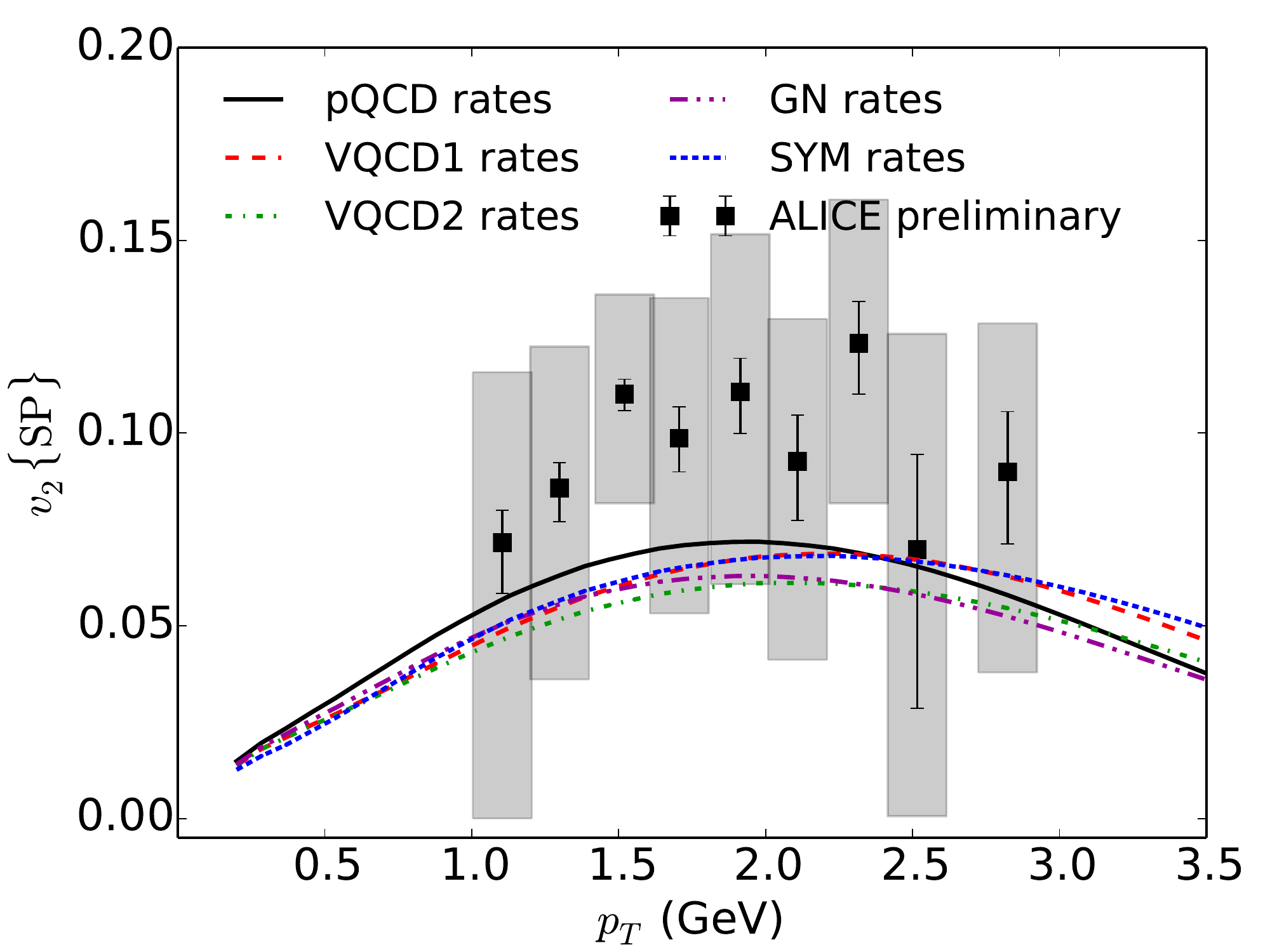}
	\caption{The result for anisotropic flow in LHC \cite{Iatrakis:2016ugz}. (Color online) Direct photon anisotropic flow $v_{2}$ in 0-40\% Pb+Pb collisions at 2.76 $A$\,TeV compared with the ALICE measurements \cite{Lohner:2012ct}. }
	\label{fig6.3}
\end{figure}

In Figs.~\ref{fig6.2} and \ref{fig6.3}, direct photon anisotropic flow coefficients, $v_{2,3}\{\mathrm{SP}\}(p_T)$(with scalar-product method), are shown at the RHIC and LHC energies together with the experimental measurements. Since the underlying hydrodynamic medium is kept fixed for all sets of calculations, we here show curves without statistical error bands for better visual comparisons. At both collision energies, the weakly-coupled QCD rates gives the largest direct photon $v_n$ in the intermediate $p_T$ region, $1 < p_T < 2.5$ GeV. At the higher $p_T > 3.0$ GeV, the $v_n$ results using holographic rates are larger. This interesting hierarchy of direct photon $v_{2,3}$ is a results of the interplay between the temperature dependence of emission rates and the space-time structure of the hydrodynamic flow distribution.


\section{Small Collision Systems}
\begin{figure}[h!]
	\centering
	\begin{tabular}{cc}
		\includegraphics[width=0.48\linewidth]{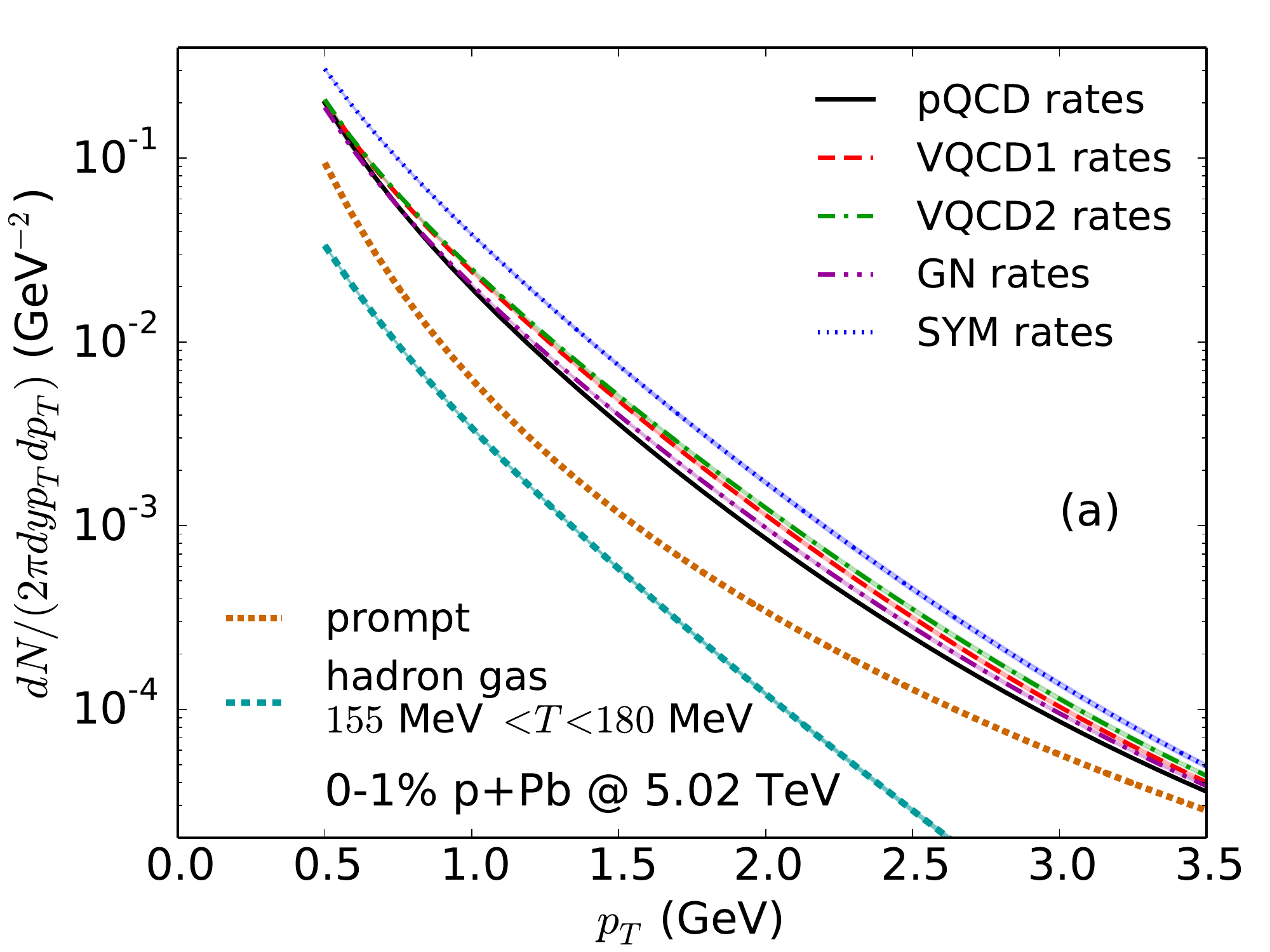} &
		\includegraphics[width=0.48\linewidth]{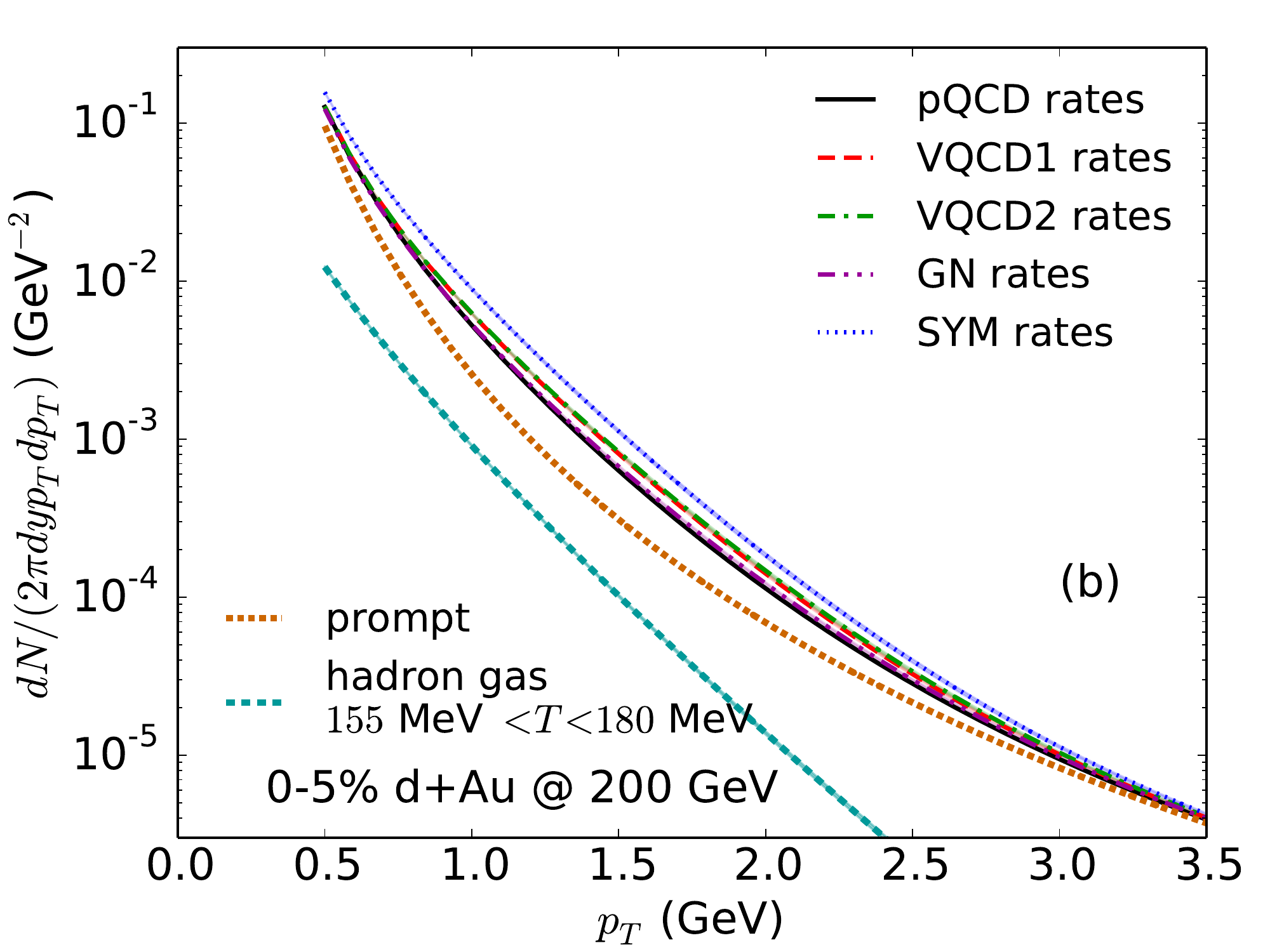} \\
		\includegraphics[width=0.48\linewidth]{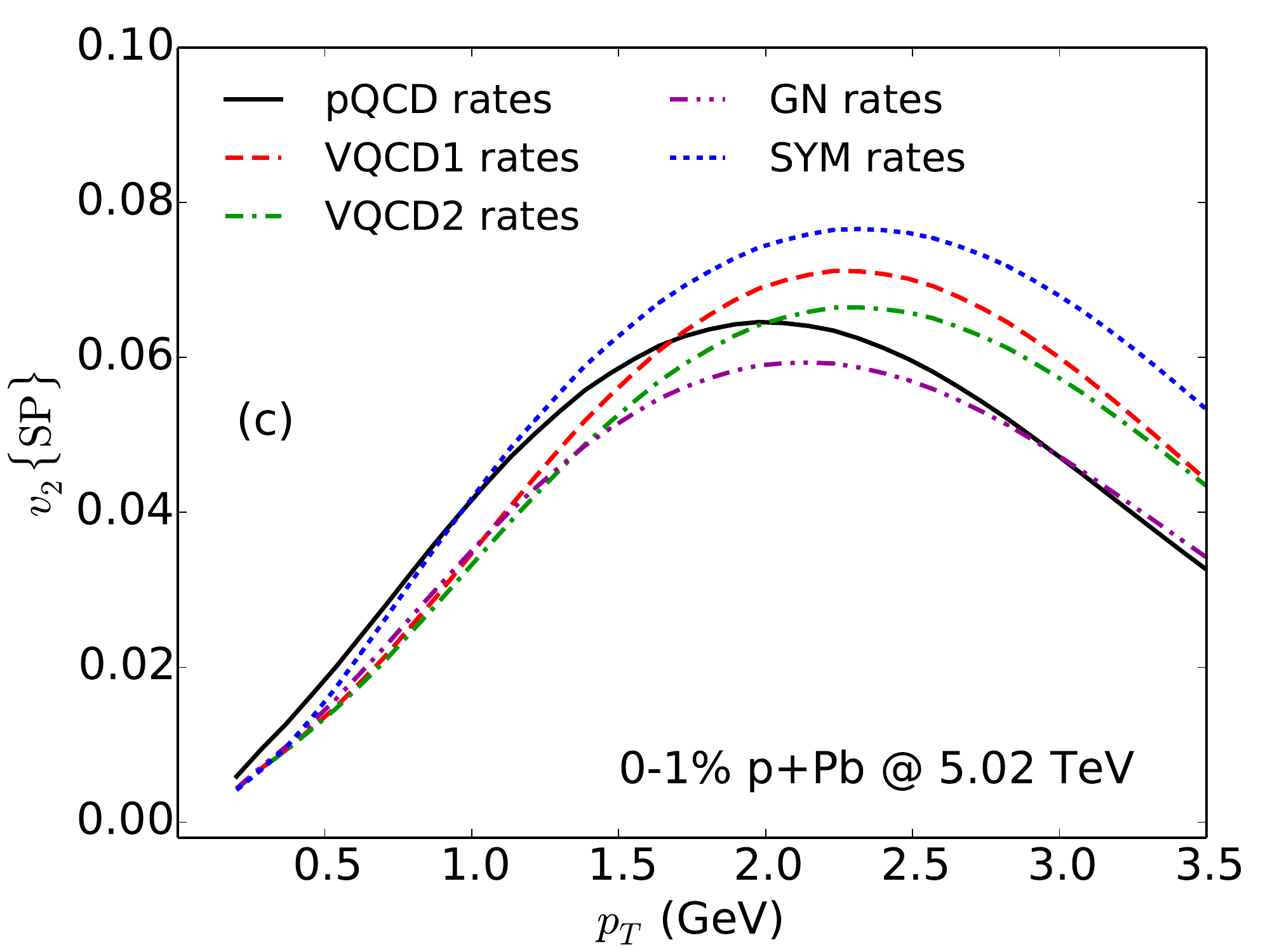} &
		\includegraphics[width=0.48\linewidth]{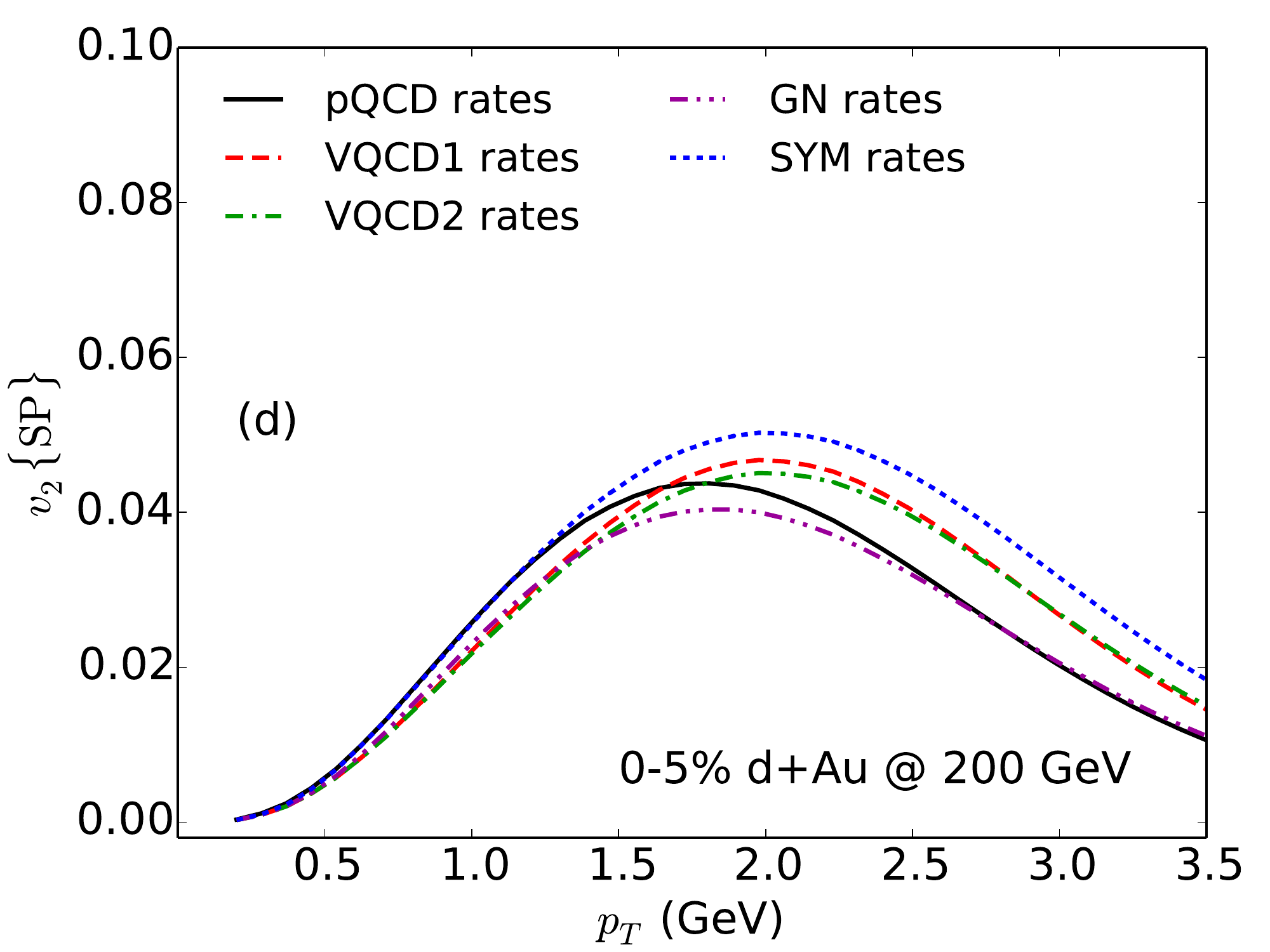} \\
		\includegraphics[width=0.48\linewidth]{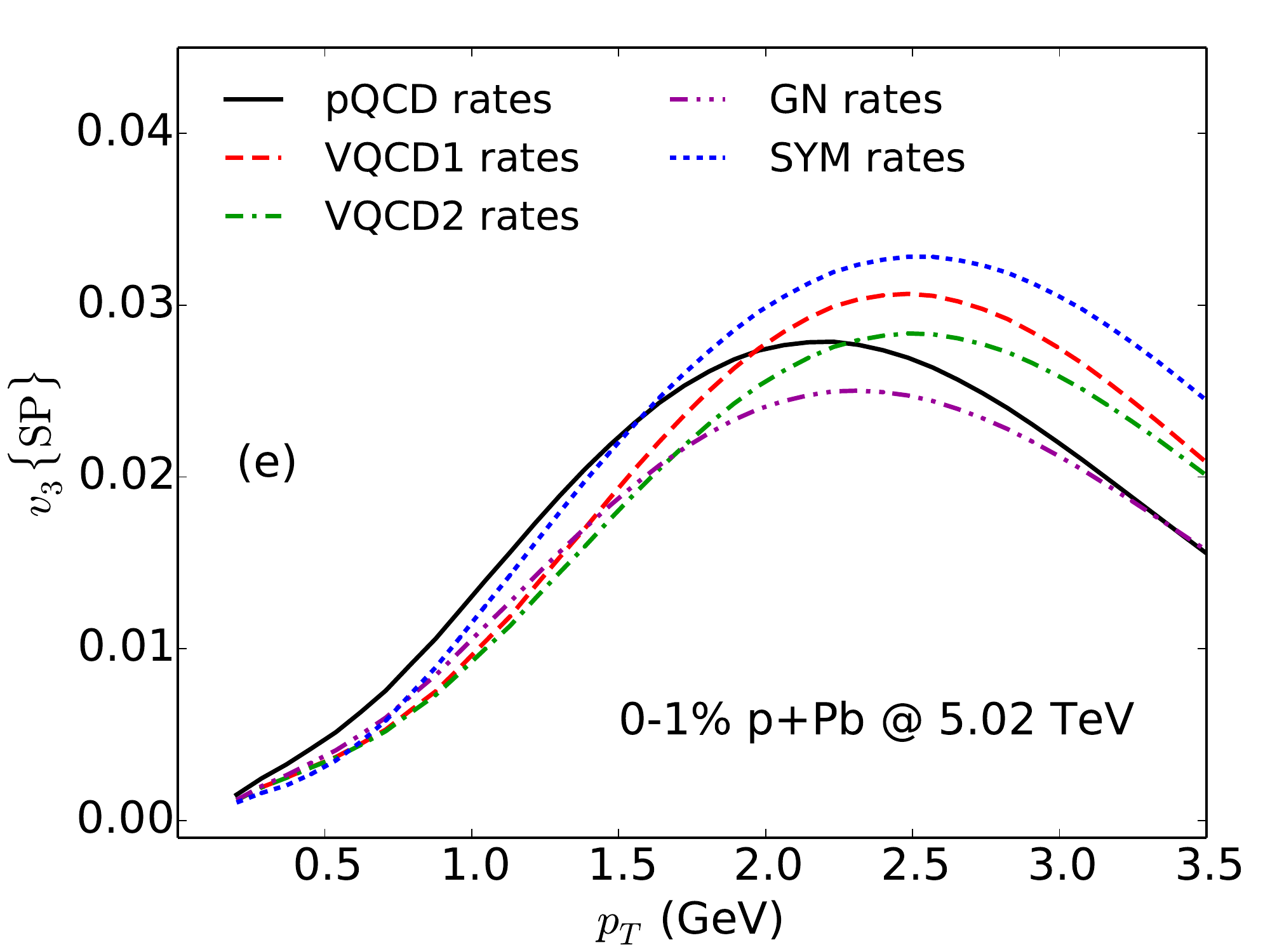} &
		\includegraphics[width=0.48\linewidth]{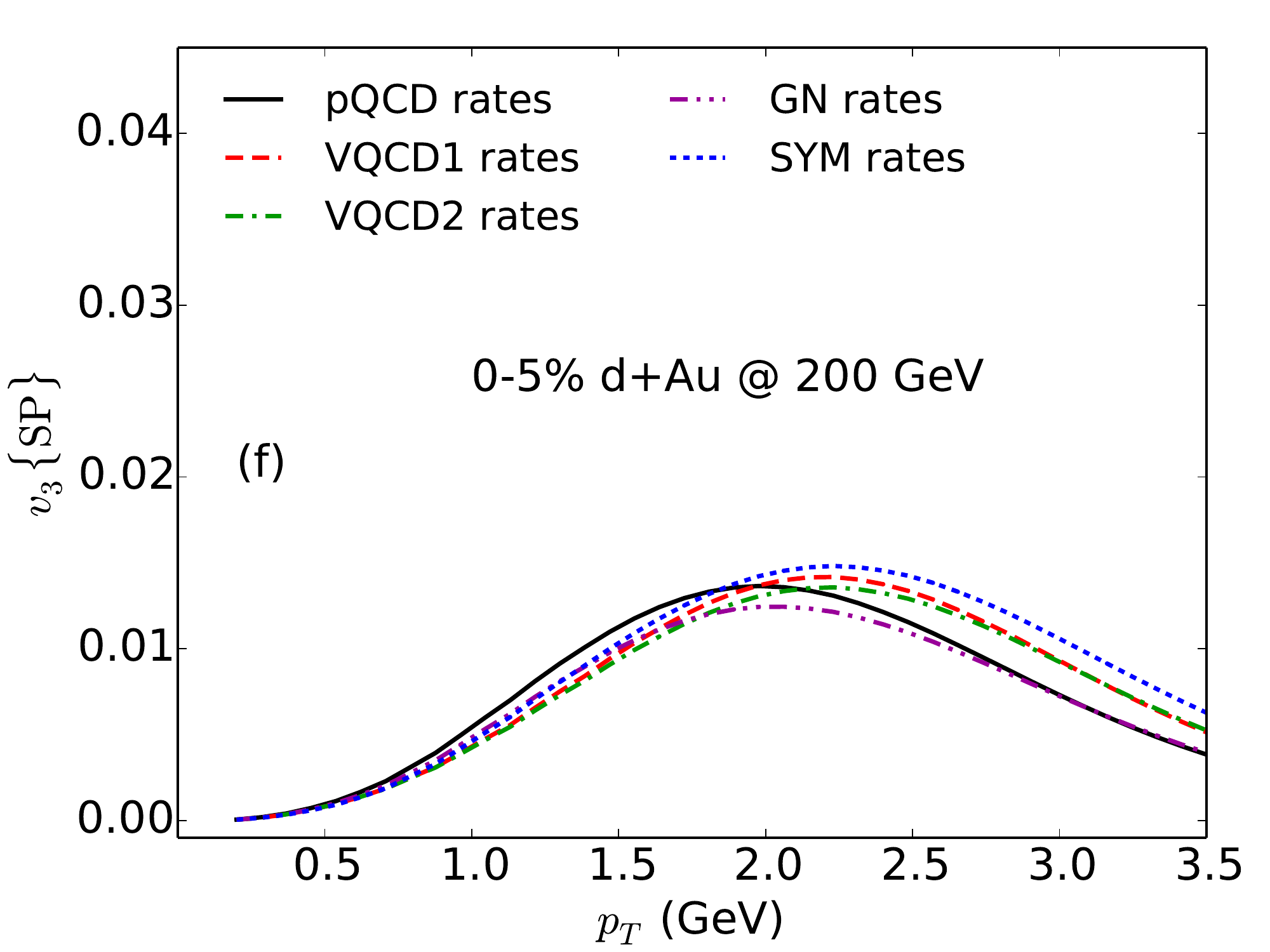}
	\end{tabular}
	\caption{Results for small collision systems \cite{Iatrakis:2016ugz}. (Color online) Direct photon spectra from 0-1\% p+Pb collisions at 5.02 TeV $(a)$ and 0-5\% d+Au collisions at 200 GeV $(b)$ using different sets of emission rate in the QGP phase. Direct photon anisotropic flow coefficients, $v_{2,3}(p_T)$, in 0-1\% p+Pb collisions at 5.02 TeV $(c,e)$ and 0-5\% d+Au collisions at 200 GeV $(d,f)$ using different sets of emission rate in the QGP phase.}
	\label{fig6.4}
\end{figure}
Recently, sizable thermal radiation was found in high multiplicity light-heavy collisions in Ref.~\cite{Shen:2015qba}. Owing to large pressure gradients, small collision systems, such as p+Pb and d+Au collisions, expand more rapidly compared to the larger Au+Au collisions, which yield higher freeze-out temperature. This leads to a smaller hadronic phase in these collision systems. Most of the thermal photons come from the hot QGP phase, $T > 180$ MeV \cite{Shen:2015qba}. Hence, the difference between the QGP photon emission rates should be more distinctive in these small collision systems. In \cite{Iatrakis:2016ugz}, the calibrated hydrodynamic medium based on \cite{Shen:2016zpp} is applied to study the sensitivity of direct photon observables in small systems to the different sets of QGP photon emission rates.

In Figs.~\ref{fig6.4}, direct photon spectra and their anisotropic coefficients are shown for top 0-1\% p+Pb collisions at 5.02 TeV and 0-5\% d+Au collisions at 200 GeV. In high-multiplicity events of small collision systems, thermal radiation can reach up to a factor of 2 of the prompt contribution. Similar to nucleus-nucleus collisions, the direct photon spectra using the emission rates that are derived from strongly coupled theory are larger than the QCD rates. 

The difference is smaller in the d+Au collisions compared to p+Pb collisions at the higher collision energy. Although the hierarchy of direct photon anisotropic flow coefficients remains the same as those in nucleus-nucleus collisions, the splittings among the results using different emission rates are larger in 0-1\% p+Pb collisions at 5.02 TeV. The direct photon anisotropic flow coefficients in small collision systems show a strong sensitivity to the QGP photon emission rates.

\section{Concluding Remarks}
In \cite{Iatrakis:2016ugz}, the agreement with experiments in spectra therein is improved compared with the previous study by using the pQCD rate. On the other hand, the deviation in flow is increased at low $p_T$ but decreased at high $p_T$. In small collision systems, holographic models lead to enhancements in both spectra and flow. The findings in \cite{Iatrakis:2016ugz} may emphasize the strong influence of thermal photons from the QGP phase on the direct-photon flow at high $p_T$, where hadronic contributions are highly suppressed. The enhancement of flow in this region stems from the amplification of the weight of late-time emission and the amplitude of thermal-photon emission in the QGP phase. 

On the contrary, the study may further {\em suggest that the hadronic contributions are responsible for the flow at low $p_T$.} 
In contrast to the nucleus-nucleus collisions, the dominance of QGP photons could be more pronounced in small collision systems for larger $p_T$ window. Therefore, future measurements of direct-photon spectra and flow in small systems will be crucial to understand the electromagnetic property of the sQGP.

Acknowledgments:
The work of I.I. is part of the D-ITP consortium, a program of the Netherlands Organisation for Scientific Research (NWO) that is funded by the Dutch Ministry of Education, Culture and Science. The work of E. K. was partially supported by  European Union's Seventh Framework Programme under grant agreements (FP7-REGPOT-2012-2013-1) no 316165 and the Advanced ERC grant SM-grav, No 669288. D.Y. was supported by the RIKEN Foreign Postdoctoral Researcher program. C. S. was supported by the U.S. Department of Energy, Office of Science under contract No. DE- SC0012704 and the Natural Sciences and Engineering Research Council of Canada.



\bibliographystyle{elsarticle-num}







\end{document}